# Reversible strain-induced magnetization switching in FeGa nanomagnets: Pathway to a rewritable, non-volatile, non-toggle, straintronic memory cell for extremely low energy operation

Hasnain Ahmad[1], Jayasimha Atulasimha[2] and Supriyo Bandyopadhyay[1]

[1]Department of Electrical and Computer Engineering
[2]Department of Mechanical and Nuclear Engineering
Virginia Commonwealth University
Richmond, VA 23284, USA

Reversible straintronic switching of a nanomagnet's magnetization between two stable or metastable states promises ultra-energy-efficient non-volatile memory. Here, we report strain-induced magnetization switching in ~300 nm sized FeGa nanomagnets delineated on a piezoelectric PMN-PT substrate. Voltage of one polarity applied across the substrate generates compressive strain in a nanomagnet and switches its magnetization to one state, while voltage of the opposite polarity generates tensile strain and switches the magnetization back to the original state, resulting in "non-toggle" switching. The two states can encode the two binary bits, and, using the right voltage polarity, one can write either bit deterministically.

There are two desirable features of rewritable magnetic memory technology other than the usual density, speed and endurance. First, the writing scheme should not depend on what the initial stored bit is since that would avoid having to read the initial bit before writing the new one. This is called "non-toggle" memory. Second, and perhaps more important, the writing scheme should be minimally dissipative, meaning that the magnetization switching methodology should be maximally energy-efficient.

There are many schemes for altering the magnetization state of a nanomagnet to enable the writing of a bit. The oldest is to use a local magnetic field generated by an on-chip current, which is, regrettably, extremely dissipative [1] and would dissipate at least $10^7$ kT of energy per write operation at room temperature [2]. The second is to use a spin-polarized current to deliver a spin-transfer torque [3] or induce domain wall motion [4]. These are also extremely dissipative strategies and dissipate between $10^4$ kT and $10^7$ kT of energy per write step [5, 6]. The recent use of the giant spin Hall effect to generate a spin polarized current [7] may end up reducing the energy dissipation to perhaps $10^4$ kT per step, but a much more energy-efficient approach is to use two-phase multiferroics (a magnetostrictive nanomagnet delineated on a piezoelectric film). An electrostatic potential applied across the piezoelectric film generates strain in that layer, which is partially transferred to the magnetostrictive layer and rotates the latter's magnetization via the Villari effect. According to theoretical calculations, this mode results in ~10 kT (T = 300 K) of energy dissipation at very low temperatures [8] and ~100 kT at room temperature to switch a nanomagnet's magnetization state in less than 1 ns [9].

There are a number of reports of switching the magnetization of nanoscale two-phase multiferroics with electrically generated strain [10 – 12]. There is also a report of switching the resistance of a magneto-

Corresponding author: S. Bandyopadhyay, Email: sbandy@vcu.edu

tunneling junction (MTJ) whose soft layer is a two-phase multiferroic (CoFeB/PMN-PT) with electrically generated strain [the MTJ is a complete memory element with read/write], but the MTJ had tens of μm-scale dimensions (not nanoscale) [13]. Most important, all of the above experiments employed ferromagnets with low saturation magnetostriction (Co, Ni, CoFeB), which is not conducive to energy efficiency since the stress needed to rotate the magnetization (and correspondingly the voltage needed to generate the stress) is inversely proportional to the magnetostriction coefficient of the magnetostrictive component. In this work, we have used FeGa as the magnetostrictive component of a nanoscale multiferroic and demonstrate stress-induced change in its magnetization state. Because FeGa has a much higher magnetostriction (approximately an order of magnitude higher in bulk) than Co or Ni, the electric field needed to alter its magnetization is relatively small. There are previous demonstrations of magnetization rotation in FeGa layers with strain [14-17], but the samples were, again, not nanoscale, except in [18] which did not demonstrate repeatable switching. Even more important, none of the above experiments addressed, let alone demonstrate, the *"non-toggle" behavior*, i.e., the magnetization being driven to one state with one sign of stress/strain and restored back to the other (original) state with the opposite sign of stress/strain. Here, we have achieved this feat.

Elliptical FeGa nanomagnets of major axis ~300 nm, minor axis ~240 nm, and thickness ~8 nm were fabricated on a (100)-oriented PMN-PT substrate (70% PMN and 30% PT). The substrate was first poled with an electric field of 8400 V/cm in a direction which will coincide with the major axes of the nanomagnets (the major axes of the nanomagnets were nominally parallel to each other). The poled substrate was spin-coated with two layers of PMMA (e-beam resist) of different molecular weights in order to obtain superior nanomagnet feature definition: PMMA-495 Anisol and PMMA-950 Anisol at 2500 rpm spinning rate. The resists were baked at 115 Celsius for 2 minutes and exposed in a Hitachi SU-70 SEM with a Nabity attachment using 30 kV accelerating voltage and 60 pA beam current. Subsequently, the resists were developed in MIBK:IPA (1:3) for 90 seconds followed by rinsing in cold IPA.

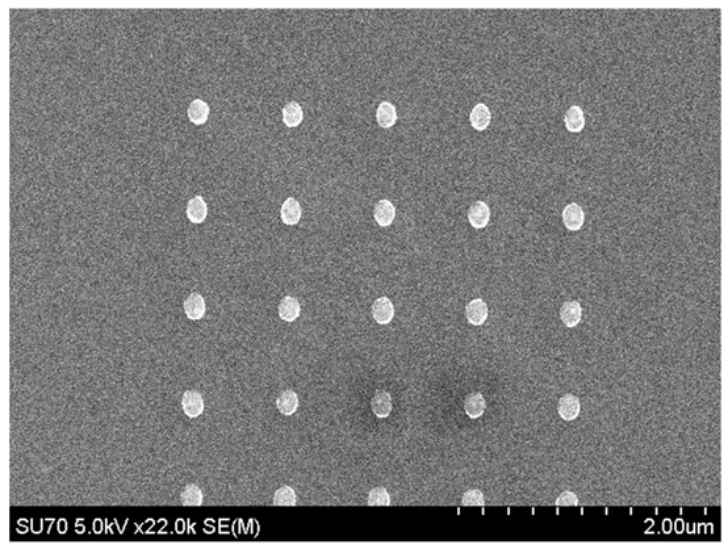

Figure 1: Scanning electron micrograph of nominally elliptical FeGa nanomagnets delineated on a silicon substrate.

Corresponding author: S. Bandyopadhyay, Email: sbandy@vcu.edu

For nanomagnet delineation, a 3-3.5 nm thick Ti adhesion layer was first deposited using e-beam evaporation at a base pressure of $(2-3) \times 10^{-7}$ Torr, followed by the deposition of 7-8 nm of FeGa (thickness verified with AFM) using DC magnetron sputtering of a FeGa target with a base pressure of $(2-3) \times 10^{-8}$ Torr and deposition pressure of 1 mTorr. The magnetron power was 45 W and the deposition was carried out for 40 seconds. The Ti layer is thin enough to not significantly impede strain transfer from PMN-PT to FeGa. Fig. 1 shows an SEM micrograph of an array of nanomagnets.

The sputtered FeGa films were characterized with x-ray diffraction (XRD) and layer-by-layer x-ray photoelectron spectroscopy (XPS) while etching off layers with an Ar ion beam. The XRD revealed that the film is polycrystalline with mostly (110) textured growth, while the XPS showed that the alloy composition was different in different layers with some incorporation of oxygen in the topmost layers due to ambient oxidation [18]. The magnetization (M-H) curves of FeGa layers were measured at 77 K and 300 K in a vibrating sample magnetometer and showed not only ferromagnetic behavior, but that the layer had in-plane magnetic anisotropy [18]. The measured in-plane coercivity was ~180 Oe and the out-of-plane coercivity was ~120 Oe. Interestingly, the M-H curves showed "shoulders" [18] indicative of the presence of more than one phase in FeGa, each with a different coercivity, as previously noted in other materials [19]. This indicated the presence of multiple energy barriers in the potential profiles of the FeGa nanomagnets which could result in the formation of metastable magnetization states. Stress could always drive a nanomagnet into such a state where it will remain after stress is withdrawn since the state is "metastable" and robust against thermal perturbations at room temperature. Metastable magnetization states could, of course, arise from other effects as well, such as due to pinning sites or irregular geometry of the nanomagnets caused by imperfect electron-beam lithography. There is a recent report of non-Joulian magnetostriction in FeGa [20] which could further complicate the nanoscale switching. In the grand scheme, none of this is important; all that matters is that stress of one sign drives the magnetization from one distinct state to another, where it stays after withdrawal of stress, and then stress of the opposite sign brings the magnetization back to the original state. This will result in a non-volatile, non-toggle, rewritable, straintronic memory.

In order to study magnetization switching, we first magnetized all FeGa nanomagnets on a PMN-PT wafer with a ~2 Tesla magnetic field directed along the nominal major axis of the nanomagnets. The magnetization states of nanomagnets were then determined with magnetic force microscopy (MFM). Care was taken to use low-moment tips in order to not perturb the magnetization states of the nanomagnets with the tip. MFM imaging showed that (see second vertical panels of Figures 2 and 3) most nanomagnets have been magnetized in the direction of the field, but some have been magnetized in an orientation that subtends a non-zero angle with the magnetizing field. This odd behavior is ascribed to the presence of spurious energy minima in the potential profile of these nanomagnets in a magnetic field that are caused by either the presence of multiple energy barriers due to irregular shapes, or multiple phases, or pinning sites. The magnetic field drives the nanomagnet to the energy minimum closest to the initial state, whose magnetization orientation may not be collinear with the field.

After the initial magnetizing, every nanomagnet is compressively stressed along its major axis by subjecting the PMN-PT substrate to a global average electric field of 4.4 kV/cm in a direction opposite to that of the initial poling. The field is generated by applying a potential of -2.2 kV along a 5 mm long substrate. This field strains the PMN-PT substrate owing to $d_{33}$ coupling. The value of $d_{33}$ measured in our

Corresponding author: S. Bandyopadhyay, Email: sbandy@vcu.edu

substrates in ref. [12] was 1000 pm/V. Therefore, the average strain generated in the PMN-PT substrate is 440 ppm. This strain is partially or completely transferred to the FeGa nanomagnets, resulting in a maximum stress of ~44 MPa in the nanomagnets since the Young's modulus of FeGa is about 100 GPa.

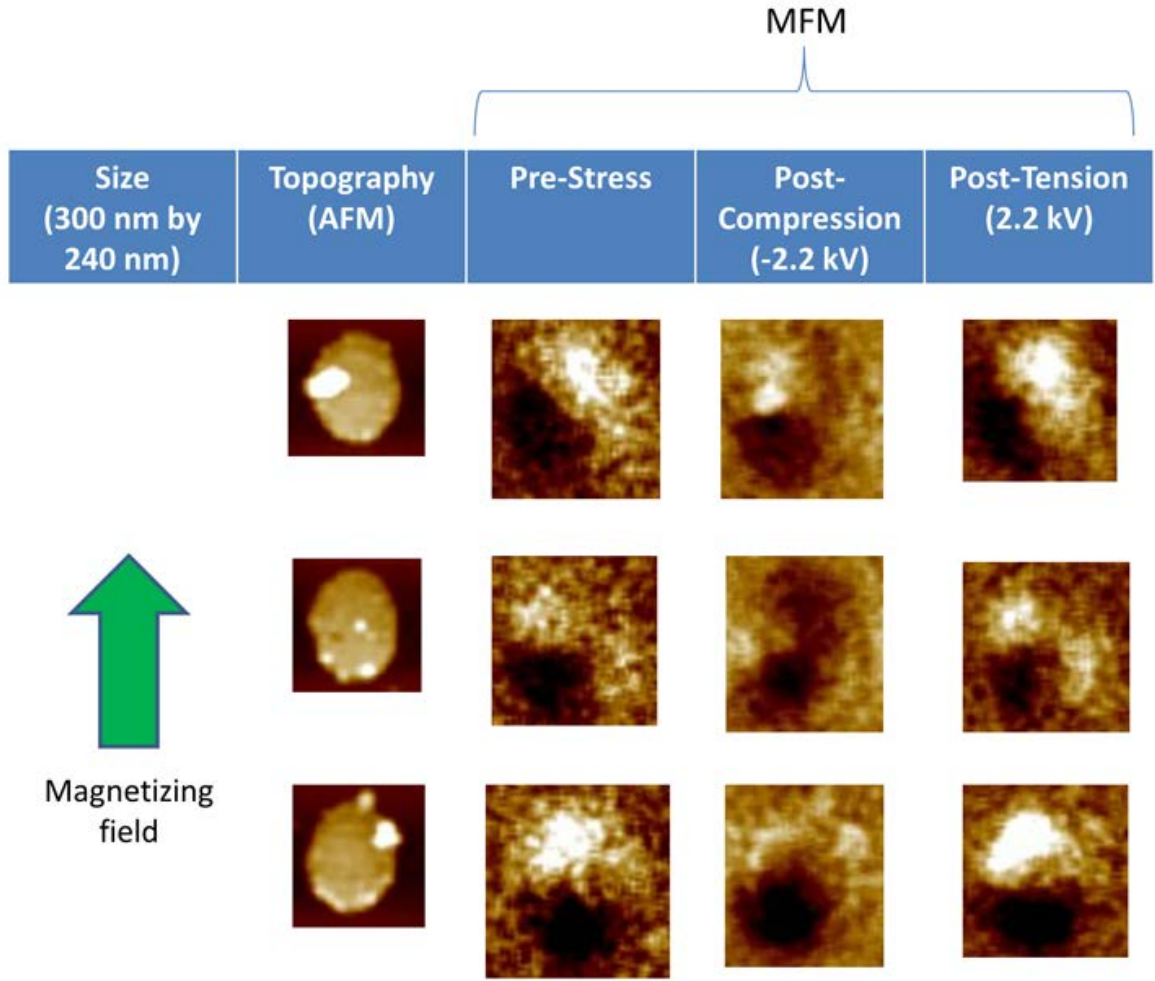

Figure 2: Magnetic force (MFM) and atomic force (AFM) micrographs of isolated elliptical FeGa nanomagnets that have been magnetized with a magnetic field in the direction indicated by the thick vertical green arrow. Starting from the left, the first vertical panel shows the AFM image, the second vertical panel shows the initial magnetization state after magnetization (note that the magnetization is not always in the direction of the magnetizing field), the third vertical panel shows the new magnetization state after compressive stress is applied and withdrawn, and the last vertical panel shows the magnetization state after tensile state is applied and withdrawn. Note that compressive stress takes the magnetization to a state different from the initial one and tensile stress brings it back to the original state.

Compressive stress makes the magnetization evolve to a new state, and the magnetization stays there after the stress (electric field) is removed, showing that the new state is "non-volatile". This is shown in the third vertical panels of Figures 2 and 3. Next, *tensile* stress is applied along the major axis of the nanomagnets by reversing the polarity of the voltage from -2.2 kV to +2.2 kV. The fourth vertical panels of Figures 2 and 3 show that all nanomagnets imaged returned to the original state after experiencing tension (it is possible, however, that some nanomagnets not imaged failed to return to their original

Corresponding author: S. Bandyopadhyay, Email: sbandy@vcu.edu

states). All this shows two important features: First, one can "rewrite" a bit in the nanomagnets after the first writing (rewritable non-volatile memory), and second, the switching is "non-toggle". Writing does not require toggling the previously stored bit; if we apply a positive voltage and tensile stress, then we will always deterministically write the bit 1 irrespective of whether the previously stored bit was 1 (no toggling required) or 0 (toggling required). The same is true if we wish to write the bit 0. In fact, we do not even need to know what the previously stored bit was, which avoids a read step.

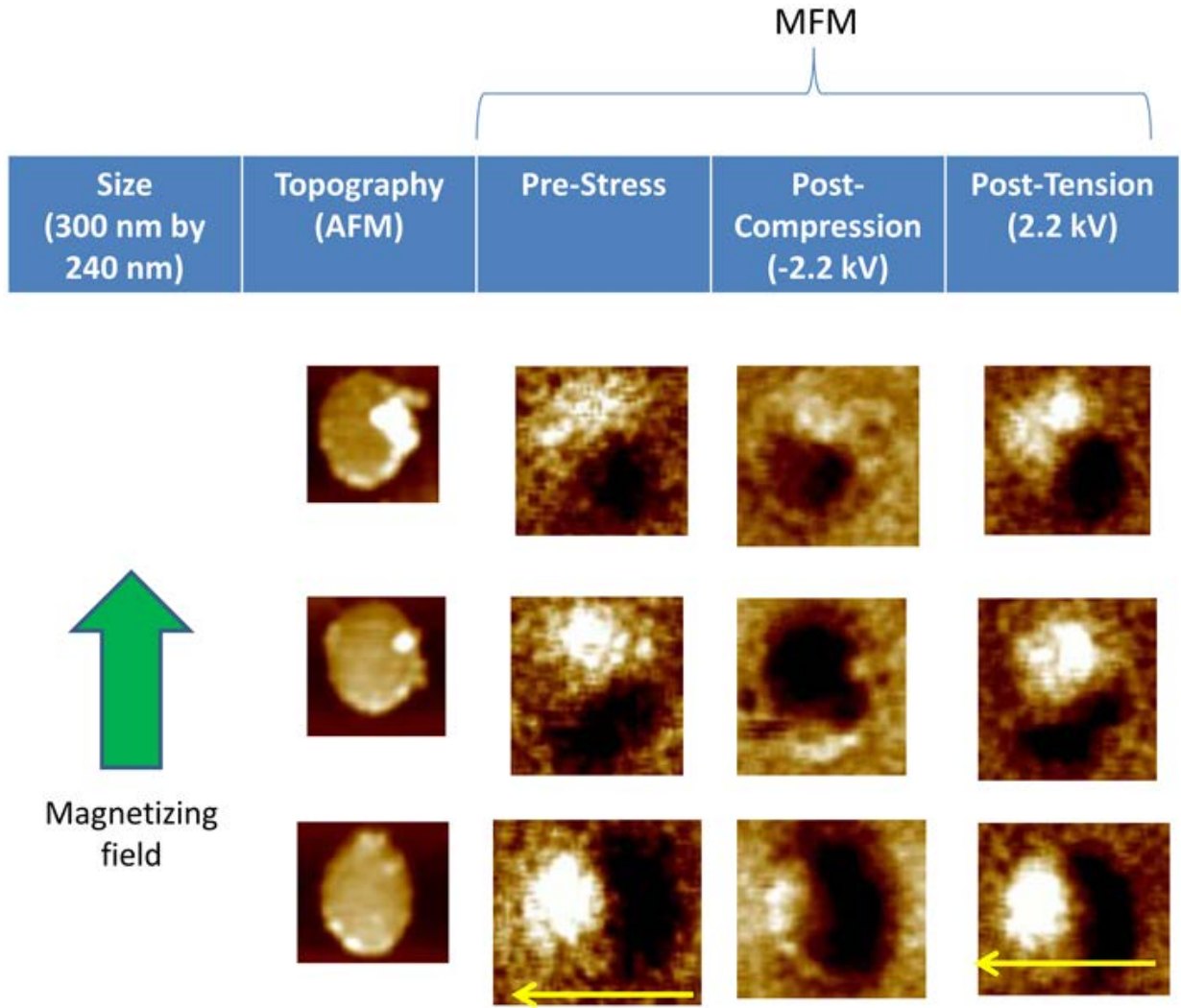

Figure 3: Same as Fig. 2, but for a different set of nanomagnets. Here, the nanomagnet in the last row was magnetized in a direction almost perpendicular to the magnetizing field showing that there is a deep energy minimum corresponding to that orientation and the nanomagnet prefers to go there even in the presence of the magnetizing field. Compressive stress, however, seems to drive it out of that state, but subsequent application of tensile stress brings it back to that state, just like in the case of the other nanomagnets.

Corresponding author: S. Bandyopadhyay, Email: sbandy@vcu.edu

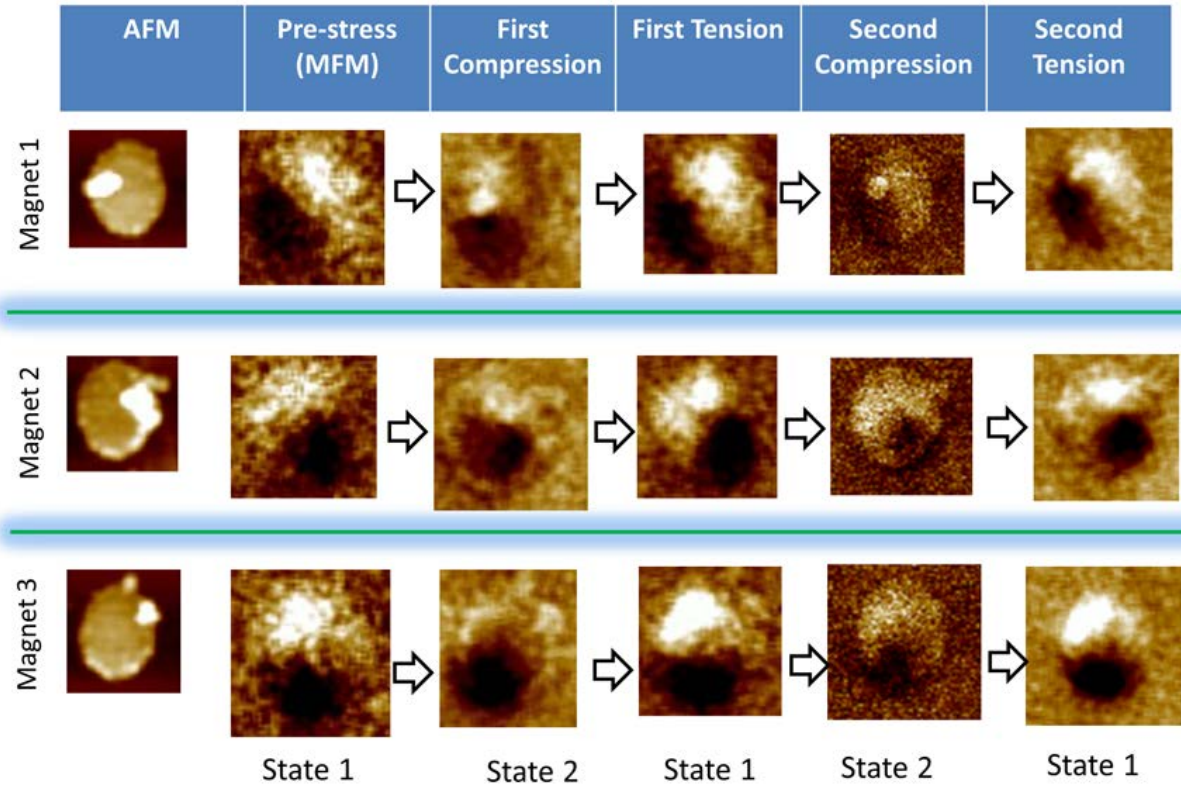

Figure 4: Magnetic force microscope images of nanomagnets showing repeatability of the switching. A nanomagnet cycles through its two magnetization states repeatedly with successive compression and tension.

In Figure 4, we show magnetic force micrographs of three FeGa nanomagnets to establish that the stress-induced magnetization switching between the two states is somewhat *repeatable*. In successive vertical panels, we present the initial magnetization state, the state after the first compression cycle, the state after the first tension cycle, the state after the second compression cycle, and the state after the second tension cycle. Compression always drives the magnetization away from its pre-stress state and tension always brings it back to the pre-stress state. In memory applications, this repeatability is important for *endurance*. We could fashion a memory element by using a magneto-tunneling junction (MTJ) with an FeGa soft layer and magnetize the hard layer permanently in a direction anti-parallel to the pre-stress magnetization orientation of the soft layer. Then, tensile stress applied with one voltage polarity will always take the MTJ resistance to a high state (encoding, say, bit 1) and compressive stress applied with the other voltage polarity will take the MTJ resistance to a different state encoding the logic complement of the bit 1.

In conclusion, we have demonstrated the core component of a *straintronic* non-volatile, rewritable, non-toggle memory element. Since the magnetization switching is induced by strain, it should be remarkably energy-efficient going by all available theoretical predictions. Here, we unfortunately had to apply a global strain with a large voltage across a large substrate (5 mm) because delineating local contact pads around each nanomagnet to apply the field locally was lithographically challenging for us. However, such lithography is not challenging for a fabrication foundry. If local strain is generated with local contacts in the manner of, say, ref. [21], then the electric field of 4.4 kV/cm had to be generated in a ~100 nm piezoelectric layer, resulting in a switching voltage *V* of only 44 mV. The scheme of [21] uses a different

Corresponding author: S. Bandyopadhyay, Email: sbandy@vcu.edu

electrode configuration, but it generates biaxial strain which is even better and may further reduce the voltage. With an electrode capacitance $C$ of ~1 fF [22], the resulting energy dissipation ~$CV^2$ to write a bit would have been only ~2 aJ (477 kT at room temperature). In contrast, present day mainstream spin-transfer-torque random access memory (STT-RAM) dissipates about $10^7$ kT of energy per write operation [6] and spin-Hall based versions will dissipate ~$10^4$ kT. Therefore, this experiment lays the foundation of a remarkably energy-efficient straintronic memory technology.

This work was supported by the US National Science Foundation under grants ECCS-1124714 and CCF-1216614. The sputtering of FeGa nanomagnets was carried out at the National Institute of Standards and Technology, Gaithersburg, Maryland, USA.

**REFERENCES**

1. M. T. Alam, M. J. Siddiq, G. H. Bernstein, M. Niemier, W. Porod and X. S. Hu, IEEE Trans. Nanotechnol., **9**, 348 (2010).
2. M. Salehi-Fashami, J. Atulasimha and S. Bandyopadhyay, Nanotechnology, **23**, 105201 (2012).
3. D. C. Ralph and M. D. Stiles, J. Magn. Magn. Mater., **320**, 1190 (2008).
4. M. Yamanouchi, D. Chiba, F. Matsukura and H. Ohno, Nature, **428**, 539 (2004)
5. S. Fukami et al., Symp. VLSI Technol., Digest of Technical Papers (IEEE), 230 (2009).
6. P. K. Amiri and K. L. Wang, Spin, **2**, 1240002 (2012).
7. L. Liu, C.-F. Pai, Y. Li, H. W. Tseng, D. C. Ralph, and R. A. Buhrman, Science, **336**, 555 (2012).
8. K. Roy, S. Bandyopadhyay and J. Atulasimha, Appl. Phys. Lett., **99**, 063108 (2011).
9. K. Roy, S. Bandyopadhyay and J. Atulasimha, J. Appl. Phys., **112**, 023914 (2012).
10. T-K Chung, S. Keller and G. P. Carman, Appl. Phys. Lett., **94**, 132501 (2009).
11. M. Buzzi, R. V. Chopdekar, J. L. Hockel, A. Bur, T. Wu, N. Pilet, P. Warnicke, G. P. Carman, L. J. Heyderman and F. Nolting, Phys. Rev. Lett., **111**, 027204 (2013).
12. N. D'Souza, M. Salehi-Fashami, S. Bandyopadhyay and J. Atulasimha, arXiv:1404.2980.
13. P. Li, A. Chen, D. Li, Y. Zhao, S. Zhang, L. Yang, Y. Liu, M. Zhu, H. Zhang and X. Han, Adv. Mater. **26**, 4320 (2014).
14. J. L. Weston, A. Butera, T. Lograsso, M. Shamsuzzoha, I. Zana, G. Zangari and J. Barnard, IEEE Trans. Magn., **38**, 2832 (2002).
15. R. R. Basantkumar, B. J. Hills Stadler, W. P. Robbins and E. M. Summers, IEEE Trans. Magn., **42**, 3102 (2006).
16. P. Zhao, Z. Zhao, D. Hunter, R. Suchoski, C. Gao, S. Matthews, M. Wuttig and I. Takeuchi, Appl. Phys. Lett., **94**, 243507 (2009).
17. T. Brintlinger, S-H. Lim, K. H. Baloch, P. Alexander, Y. Qi, J. Barry, J. Melngailis, L. Salamanca-Riba, I. Takeuchi and J. Cumings, Nano Letters,**10**, 1219 (2010).
18. H. Ahmad, J. Atulasimha and S. Bandyopadhyay, arXiv:1506.06189.
19. M. E. Jamer, B. A. Assaf, S. P. Bennett, L. H. Lewis and D. Heiman, J. Magn. Magn. Mater., **358-359**, 259 (2014).
20. H. D. Chopra and M. Wuttig, Nature, **521**, 340 (2015).
21. J. Cui, J. L. Hockel, P. K. Nordeen, D. M. Pisani, C. y. Liang, G. P. Carman, and C. S. Lynch, Appl. Phys. Lett., **103**, 232905 (2013).

Corresponding author: S. Bandyopadhyay, Email: sbandy@vcu.edu

22. A. K. Biswas, S. Bandyopadhyay and J. Atulasimha, Appl. Phys. Lett., **105**, 072408 (2014).

Corresponding author: S. Bandyopadhyay, Email: sbandy@vcu.edu